\documentstyle[12pt,epsfig]{article}

\newcommand{\npb}[3]{Nucl.~Phys.~#1 (19#2) #3}
\newcommand{\prl}[3]{Phys.~Rev.~Lett.~#1 (19#2) #3}

\newcommand{\pr}[3]{Phys.~Rev.~D#1 (19#2) #3}

\newcommand{\lsim}{\raisebox{-0.13cm}{~\shortstack{$<$ \\[-0.07cm] $\sim$}}~}
\newcommand{\gsim}{\raisebox{-0.13cm}{~\shortstack{$>$ \\[-0.07cm] $\sim$}}~}

\newcommand{\ra}{\rightarrow}
\newcommand{\ee}{e^+e^-}
\newcommand{\s}{\\ \vspace*{-3mm} }
\newcommand{\nn}{\noindent}
\newcommand{\non}{\nonumber}
\newcommand{\beq}{\begin{eqnarray}}
\newcommand{\eeq}{\end{eqnarray}}

\newcommand{\tb}{\tan\beta}

\begin{document}

\begin{titlepage}

\begin{flushright}
KA--TP--02--96\\
February 1996 \\
\end{flushright}

\def\thefootnote{\fnsymbol{footnote}}

\vspace{1cm}

\begin{center}

{\large\sc {\bf  Loop Induced Higgs Boson Pair Production}}

\vspace{.3cm}

{\large\sc {\bf  at e$^+$e$^-$ Colliders.}}

\vspace{1cm}

{\sc A.~Djouadi\footnote{Supported by Deutsche Forschungsgemeinschaft
DFG (Bonn).}, V. Driesen and C. J\"unger}

\vspace{1cm}

Institut f\"ur Theoretische Physik, Universit\"at Karlsruhe,\\
\vspace{0.2cm}
D--76128 Karlsruhe, FR Germany. \\

\end{center}

\vspace{1.6cm}

\begin{abstract}

\nn We analyse the loop induced production of Higgs boson pairs at
future high--energy $\ee$ colliders, both in the Standard Model and in
its minimal supersymmetric extension. The cross sections for Standard
Model Higgs pair production through $W/Z$ boson loops, $\ee \ra H^0
H^0$, are rather small but the process could be visible for high enough
luminosities, especially if longitudinal polarization is made available.
In the Minimal Supersymmetric Standard Model, the corresponding
processes of CP--even or CP--odd Higgs boson pair production, $\ee \ra
hh, HH, Hh$ and $\ee \ra AA$ have smaller cross sections, in
general. The additional contributions from chargino/neutralino and
slepton loops are at the level of a few percent in most of the
supersymmetric parameter space.

\end{abstract}

\end{titlepage}

\def\thefootnote{\arabic{footnote}}
\setcounter{footnote}{0}
\setcounter{page}{2}

\subsection*{1. Introduction}

The search for scalar Higgs particles and the exploration of the
electroweak gauge symmetry breaking sector will be one of the main goals of
future high--energy colliders. While Higgs particles will probably be
first produced at the Large Hadron Collider LHC, the clean environment
of high--energy $\ee$ linear colliders would be required for a detailed
investigation and for the high--precision measurement of the fundamental
properties of the Higgs particles \cite{R1}, a crucial requirement to
establish the Higgs mechanism as the basic mechanism to generate the
masses of the known particles. To this end, a precise calculation of the
branching ratios of all important decay channels as well as the cross
sections of the production mechanisms, is mandatory \cite{R1,R2}. \s 

The main production mechanisms of the Standard Model (SM) Higgs particle
$H^0$ at high--energy $\ee$ colliders \cite{R2}, are the bremsstrahlung
process \cite{R3}, $\ee \ra H^0Z$, and the vector boson fusion
mechanisms \cite{R4}, $\ee \ra W^* W^*/Z^* Z^* \ra H^0 \nu\bar{\nu}/ H^0
\ee$. Besides providing the experimental signals for the Higgs
particles, these processes would also allow to measure the Higgs
couplings to gauge bosons, and to test that they are indeed proportional 
to the $W/Z$ masses, a fundamental prediction of the Higgs mechanism.
 Some of the couplings of Higgs particles can also be measured by
considering the decay branching ratios \cite{R5} or higher--order
production processes. For instance, the Higgs boson self--coupling can
be accessible in the pair production processes $\ee \ra H^0 H^0 Z$
and/or $W^* W^*/Z^*Z^* \ra H^0 H^0$ \cite{R6}, while the Yukawa coupling
of light Higgs bosons to top quarks can be directly measured in 
the process $\ee \ra t\overline{t}H^0$ \cite{R7}. \s 

Higher order processes might also be useful to discriminate between the
Higgs sector of the Standard Model from the usually more complicated
scalar sectors of its possible extensions. Among these extensions,
supersymmetric theories and in particular the Minimal Supersymmetric
Standard Model (MSSM) are the most natural, theoretically. In the MSSM,
the Higgs sector is enlarged to contain two scalar doublet
fields leading to a spectrum of five Higgs particles: two neutral
CP--even ($h$ and $H$), a CP--odd ($A$) and two charged ($H^\pm$) Higgs
bosons \cite{R1}. 
In the decoupling regime \cite{R8}, where the $A,H$ and $H^\pm$ bosons are
very heavy, the lightest Higgs boson $h$ has exactly the same
properties as the SM Higgs boson, except that its mass is restricted to
be smaller than $M_h \lsim 140$ GeV. \s 

If the genuine supersymmetric particles were too heavy to be
kinematically accessible in collider experiments, the only way to
distinguish between the SM and the lightest MSSM Higgs boson in this
limit, is to search for loop induced contributions of the supersymmetric
particles, which could give rise to sizeable deviations from the
predictions of the SM. Well known examples of this loop induced
processes are the $\gamma \gamma$ widths of the Higgs
particles [which can be measured in the decay Higgs $\ra \gamma \gamma$
at hadron colliders, and more precisely at high--energy $\ee$ colliders
in the direct production of Higgs particles via laser--photon fusion,
$\gamma \gamma \ra$ Higgs] or the process $\ee \ra Z$+Higgs  which in
the MSSM receive extra contributions from supersymmetric gaugino and
sfermion loops \cite{R10,R10a,R10b}. \s 

Another type of such discriminating processes is the pair production of
Higgs bosons which will be analyzed in this paper. In the Standard
Model, where it has been first discussed in Ref.\cite{R11}, the process 
\beq
\ee \ra H^0H^0
\eeq
is mediated only by $W$ and $Z$ boson loops, Fig.1a, while in the
Minimal Supersymmetric extension, additional contributions to the
corresponding process 
\beq
\ee \ra hh
\eeq
will originate from chargino, neutralino, selectron and sneutrino loops,
as well as loops built up by the associated $A$ and $H^{\pm}$ bosons; Fig.1b. 
At high center--of--mass energies, $\sqrt{s}
\gsim 1$ TeV, the cross sections of the processes are rather small, 
being of the order of a fraction of a femtobarn. However, with the large
luminosities expected at these colliders, $\int {\cal L} \gsim 100$
fb$^{-1}$, and with the availability of longitudinal polarization of the
electron and positron beams which could increase the cross sections by
a factor of four, a few hundred events might eventually be collected in
the course of a few years, allowing for the experimental study of this
final state\footnote{This process is similar 
to the double Higgs production $W^*W^* \ra H^0H^0$ or $e^+e^- \ra ZH^0H^0 \ra 
\nu\overline{\nu}H^0H^0$, which consist of the same final state 
[although with missing energy] and which can be used to measure the 
trilinear Higgs self--coupling \cite{R6}.}.  \s 

The supersymmetric contributions to the process $\ee \ra hh$ in the MSSM
turn out to be rather small compared to the dominant $W/Z$ [and $A/H^\pm$]
contributions. Except in some regions of the SUSY parameter space where
they can reach the level of $\sim 15\%$, they are typically of the order
of a few percent, rendering the distinction between the SM and the light
MSSM Higgs bosons rather difficult in the decoupling regime. 
For completeness, we have also
analyzed the pair production of the heavy neutral Higgs particles of the
MSSM. Up to phase space suppression factors, the cross sections for the
processes 
\beq
\ee \ra HH \ \ \mbox{\rm and} \ \ AA 
\eeq
are of the same order as the SM Higgs pair production cross section. In
contrast, due to mixing angle factors suppression, the cross section for
\beq
\ee \ra Hh 
\eeq
is at least one order of magnitude smaller, but the supersymmetric
contributions are of the same size as the $W/Z/$Higgs contributions. The
processes $\ee \ra HA/hA$ and $\ee \ra H^+H^-$ appear already at the
tree level, and will not be considered here\footnote{The radiative
corrections to these processes, including the contributions of the
supersymmetric particles, have been recently discussed in
Refs.~\cite{R10,R10a,R12}.}. \s 

The rest of the paper is organised as follows. In the next section we
present the cross sections for the Higgs pair production process 
in the Standard Model. In section 3, we discuss the case of 
the Higgs pair production process 
in the MSSM. Some conclusions will be drawn in section 4. In the Appendix, 
we will give the lengthy analytical expressions of the cross sections in the 
MSSM.

\subsection*{2. SM Higgs Pair Production}

In the Standard Model, many Feynman diagrams could in principle
contribute to the amplitude for the Higgs boson pair production process,
eq.(1). However, in the limit of vanishing electron mass, which implies
a vanishing $Hee$ coupling, the contributions of the tree--level
diagrams where the Higgs bosons are emitted from the electron lines are
negligible. Non--zero contributions to the process $\ee \ra H^0 H^0$ can
therefore only come from one--loop diagrams. \s 

Because of the exact chiral symmetry for $m_e=0$, all diagrams involving
the one--loop $H\ee$ vertex must give zero contributions [in fact this
statement is valid to all orders in perturbation theory]. Furthermore,
because of CP invariance, the amplitudes of the diagrams with $\gamma$
and $Z$ boson $s$--channel exchanges, which give rise to two Higgs
bosons through vertex diagrams, also vanish [the contribution of the
longitudinal component of the $Z$ boson will be proportional to the
electron mass, and is therefore negligible]. Additional contributions
from vertex diagrams involving the quartic $WWH^0H^0/ZZH^0H^0$ couplings
are proportional to the electron/neutrino masses and are again
negligible. The only contribution to Higgs pair production in the
Standard Model will therefore come from $W$ and $Z$ box diagrams,
Fig.1a. \s 

Working in the Feynman gauge, where the $W/Z$ boson contributions can be
split into the $g_{\mu \nu}$ and the neutral Goldstone parts, the
ultraviolet--finite amplitudes have been reduced from a complicated
tensorial form to scalar Passarino--Veltman functions \cite{R13} using
the package FORM \cite{R14}. Before the reduction to scalar functions,
we have verified that our expressions agree with those obtained in
Ref.\cite{R11}. The amplitudes have then been symmetrized and squared; a
factor of $1/2$ has been included because of the two identical particles
in the final state. Allowing for the polarization of both the electron
and positron beams, which are respectively denoted by $\lambda_-$ and
$\lambda_+$, the differential cross section of the process (1) [in the
limit $m_e =0$] is then given by 
\begin{eqnarray}
\frac{\mbox{\rm d} \sigma}{\mbox{\rm d}\cos \theta } 
= \frac{G_F^4 M_W^8} {1024\, \pi^5}\,\frac{\sqrt{1-4M_H^2/s}}
 {s(u\,t-M_H^4)} 
\hspace*{-2mm} & \Bigg\{ & \hspace*{-2mm} 
(1+\lambda_-)(1-\lambda_+) \, \Big| F^W \,+\, 
\frac{(2\,s_W^2-1)^2}{2c_W^4} \,F^Z \Big|^2  \non \\
&+ &         (1-\lambda_-)(1+\lambda_+) \, 4\, \frac{s_W^8}{c_W^8} \, \Big| 
F^Z \Big|^2 \Bigg\}
\end{eqnarray}
with $s, t$ and $u$ the usual Mandelstam variables, $\theta$ the
angle between the initial electron and one of the Higgs bosons, and
$s_W^2=1-c_W^2= \sin^2\theta_W$. The form factor $F^V$ with $V=W/Z$ can
be split into two terms corresponding to the contributions of the
$g_{\mu \nu}$ and Goldstone parts 
\beq
F^V= 2M_V^2 F^V_1 +F^V_2
\eeq
where, in terms of the scalar two, three and four--point
Passarino--Veltman functions\footnote{The complete analytical
expressions of the scalar functions can be found in Ref.\cite{R15} for
instance; for their numerical evaluation we have used the package FF
\cite{R16}.}, denoted respectively by $B_0, C_0$ and $D_0$, the form
factors $F^V_{1,2}$ are given by 
\begin{eqnarray}
F_1^V 
&=& \Bigg\{ \bigg[
      2\,M_H^2\,u-M_V^2\,(u-t)-u\,(u+t) \bigg] \; D_0 
     -s\;C_0\left(M_V^2,0,M_V^2,0,0,s\right) \non \\
& &  +(M_H^2-u)\bigg[ C_0\left( M_V^2,0,M_V^2,0,u,M_H^2\right)
                      +C_0\left( 0,M_V^2,M_V^2,0,M_H^2,u\right)
                \bigg] \non \\
& &        +(u-t)\;C_0\left( M_V^2,M_V^2,M_V^2,s,M_H^2,M_H^2\right)
                       \; \Bigg\} \ \ + \ 
                  \Bigg\{ t \longleftrightarrow u \Bigg\} \\
F_2^V
&= &  \Bigg\{ \left[t\,u-u^2+2\,u\,M_H^2-2\,M_H^4\right] \,
                       C_0\left(M_V^2,0,M_V^2,0,0,s\right) \non \\
& & +     u\,\left(u-M_H^2\right)\bigg[ C_0\left( M_V^2,0,M_V^2,0,u,M_H^2\right)
                                    +C_0\left( 0,M_V^2,M_V^2,0,M_H^2,u\right)
                             \bigg] \non \\
& &  -      \frac{t^2\,u - u^2\,t - 2 u^3+4\,M_H^2\,(t\,u-u^2) + M_H^4\,(t+u)}
           {2\,(s-4\,M_H^2)}\;C_0\left( M_V^2,M_V^2,M_V^2,s,
                                          M_H^2,M_H^2\right) \non \\
& &  +
      \Big[ u^2\,(u-t)+M_V^2\,u\,(t+u)+2\,u\,M_H^2\,(M_H^2-u)-2\,M_H^4\,M_V^2
      \Big]D_0 \non \\ 
& &   + \frac{ut-M_H^4}{s-4\,M_H^2}
               \Big[B_0\left(M_V^2,M_V^2,M_H^2\right) -
                     B_0\left(M_V^2,M_V^2,s\right) \Big] \Bigg\} \ \ + \ 
                  \Bigg\{ t \longleftrightarrow u \Bigg\}
\end{eqnarray}
with 
\beq
D_0\equiv D_0(M_V^2,0,M_V^2,M_V^2,0,0,M_H^2,M_H^2,s,u)  
\eeq

\bigskip 

The cross sections are shown in Fig.2a as a function of the Higgs boson
mass for two center--of--mass energies, $\sqrt{s}=500$ GeV and 1.5 TeV.
Except when approaching the $2M_H$ threshold [and the small dip near the
$WW$ threshold], the cross sections are practically constant for a given
value of the c.m. energy, and amount to $\sigma \sim 0.2$ fb at
$\sqrt{s}=500$ GeV in the unpolarized case. The decrease of the cross
sections with increasing center--of--mass energy is very mild: at $
\sqrt{s}=1.5$ TeV, the cross section is still at the level of $\sigma \sim
0.15$ fb for Higgs boson masses less than $M_H \lsim 350$ GeV. \s

The effect of polarizing the initial electron/positron beams is also
shown in Fig.2a. With left--handed polarized electrons, the cross
section $e_L^- e^+ \ra H^0 H^0$ is larger by a factor of two, while for
left--handed electrons and right--handed positrons, the cross section
$e_L^- e_R^+ \ra H^0 H^0$ is larger by a factor of four,  compared to
the unpolarized case. [These simple factors of 2 and 4 are due to the
fact that the contribution of the $W$ boson is much more important than
the one of the $Z$ boson, a mere consequence of the larger charged
current couplings compared to the neutral current couplings.] Therefore,
the availability of longitudinal polarization of the initial beams is
very important. \s 

With integrated luminosities of the order of $\int {\cal L} \sim 100$
fb$^{-1}$ which are expected to be available for future high--energy
linear colliders, one could expect a few hundred events in the course of
a few years, if both initial beams can be longitudinally polarized. \\

Fig.2b shows the angular distributions d$\sigma/$d$\cos\theta$ at a
center--of--mass energy of $\sqrt{s}=500$ GeV for a Higgs boson mass of
$M_H=150$ GeV. It is forward--backward symmetric, a consequence of the
two identical particles in the final state, and follows approximately
the d$\sigma/$d$\cos\theta \sim \sin^2 \theta$ law. Again, the angular
distribution does not strongly depend on the Higgs mass and on the c.m.
energy, except near the $2M_H$ production threshold. \s 

For Higgs masses below $M_H \lsim 140$ GeV, the signal will mainly
consist of four $b$ quarks in the final state, 
\beq
\ee \ \ra \ H^0 H^0 \ \ra \ b \bar{b} \ b \bar{b}
\eeq
since the dominant
decay mode of the SM Higgs boson in this mass range is $H^0 \ra
b\bar{b}$. 
This calls for very efficient $\mu$--vertex detectors to tag
the $b$ quark jets. Since these rare events will be searched for only
after the discovery of the Higgs boson in the main production processes,
the Higgs boson mass will be precisely known and the two mass
constraints $m(b\bar{b})=M_H$, together with the rather large number of
$b$ quarks in the final state, give a reasonable hope to experimentally
isolate the signals despite of the low rates. \s

For larger Higgs masses, $M_H \gsim 140$ GeV, since $H^0 \ra W^+W^-$ 
and $H^0 \ra ZZ$ will be the dominant decay modes of the Higgs boson, the 
signals will consist of four gauge bosons in the final state, 
\beq
\ee \ \ra \ H^0 H^0 \ \ra \ W^+W^- W^+W^- \ , \
W^+W^- Z Z \ , \
ZZZZ 
\eeq
leading to eight final fermions. These rather spectacular events should
also help to experimentally isolate the signal. \s 

Although the discussion of the background events is beyond the scope of
this paper, a few comments are nevertheless in order. \s

$(i)$ For light Higgs bosons, assuming that $b$ quarks will be
efficiently tagged, the main backgrounds  will consist of the QCD 4
$b$--jet process, $ZH^0$ and $ZZ$ pair production with the $Z$ bosons
decaying into $b\bar{b}$ pairs, the latter background being the most 
dangerous especially if $M_H \sim M_Z$. However, since $ZZ$ production
is mediated by $t$--channel electron exchange, the cross section is
strongly peaked in the forward and backward directions contrary to the
signal process as shown in Fig.2b; a strong cut on $\cos\theta$ should
considerably reduce the background, while leaving the signal almost
unaltered. \s

$(ii)$ For higher Higgs boson masses, multiple vector boson production 
will be the main background; however, since these are higher--order processes 
in the electroweak coupling, the cross sections should be small enough for 
this background to be manageable. \s

$(iii)$ Finally, one should note that double Higgs production in $WW$
fusion, $\ee \ra H^0 H^0 \nu \bar{\nu}$ and in the bremsstrahlung process
$\ee \ra H^0 H^0 + Z[\ra \nu\bar{\nu}]$ would also act as backgrounds
since the two neutrinos will escape undetected.  However, the
requirement of no missing energy in the final state will easily discard
these events. 

\subsection*{3. MSSM Higgs Pair Production}

In the case of the pair production of the CP--even Higgs bosons of the
Minimal Supersymmetric Standard Model, $\ee \ra \Phi_1 \Phi_2$ with
$\Phi_1, \Phi_2 \equiv h,H$, several additional Feynman diagrams will
contribute to the processes; Fig.1b. Besides the $W$ and $Z$ boson box
diagrams, one first has the box diagrams with the exchange of the
pseudoscalar and the charged Higgs bosons, $A$ and $H^\pm$. Then, one
has two classes of box diagrams built up by chargino/sneutrino and
neutralino/selectron loops\footnote{Vertex diagrams involving the
quartic slepton--slepton--$\Phi_1$--$\Phi_2$ couplings will give rise to
contributions which are proportional to the mixing between the two
slepton eigenstates. Since this mixing is proportional to the partner
lepton masses  $m_e$ and $m_\nu$, these diagrams give negligible
contributions.}. These two sets consist of diagrams where both Higgs
bosons couple to the neutralino/chargino or to slepton pairs, and diagrams
where one Higgs boson couples to neutralinos/charginos and the other to
sleptons. One has also to include the crossed diagrams in Fig.1b, with the
exchange of $\Phi_1$ and $\Phi_2$. \s 

For the pair production of the CP--odd Higgs boson in the MSSM, $\ee \ra
AA$, only a small subset of the previous Feynman diagrams contributes;
Fig.1c and the corresponding crossed diagrams. Indeed, because of CP
invariance the pseudoscalar Higgs boson does not couple to vector boson
pairs and to slepton pairs [in the latter case, the couplings would be
proportional to the lepton masses, $m_e$ and $m_\nu$]. The only relevant
contributions will therefore come from the diagrams with $h,H,H^\pm$
boson exchange, and those with chargino/sneutrino and
neutralino/selectron exchanges, where both $A$ bosons are emitted form
the neutralino/chargino lines. \s 

We have calculated the cross sections of the four processes eqs.(2,3)
and (4) in the MSSM, following the same steps as those discussed in the
case of the SM Higgs pair production. The analytical expressions of the
cross sections are much more involved than in the SM case, a consequence
of the many additional contributions, and in the case of $\ee \ra
Hh$, of the two different particles in the final state. These
expressions will therefore be given in the Appendix. Here, we will simply
describe our inputs and discuss the numerical results which have been
obtained.  \s 

Besides the four masses $M_h, M_H, M_A$ and $M_{H^\pm}$ of the Higgs
particles, two additional parameters determine the Higgs sector of the
MSSM at the tree level: the ratio $\tb=v_2/v_1$ of the vacuum
expectation values of the two Higgs doublet fields and a mixing angle
$\alpha$ in the neutral CP--even sector. However, supersymmetry leads to
several relations among these parameters and only two of them are in
fact independent: if the pseudoscalar Higgs mass $M_A$ and $\tb$ are
specified, all other masses and the mixing angle $\alpha$ can be derived
at the tree--level \cite{R1}. Supersymmetry imposes a strong 
hierarchical structure on the mass spectrum, $M_h<M_A<M_H$, $M_W<
M_{H^\pm}$ and $M_h<M_Z$, which however is broken by radiative
corrections \cite{R17}. \s 

The leading part of the  radiative corrections grows as the fourth power
of the top quark mass $m_t$ and the logarithm of the common squark mass
$M_S$ \cite{R17}. At the subleading level the radiative corrections will
introduce the supersymmetric Higgs mass parameter $\mu$ and the
parameters $A_t,A_b$ in the soft symmetry breaking interaction
\cite{R18}. In our analysis, we will take into account the full one--loop
corrections for the Higgs boson masses \cite{R10}; 
for the mixing angle we
will use the one--loop improved \cite{R20} relation [$\tb$ and $M_A$ are
the input parameters and $M_h$ is the full one--loop corrected $h$ boson
mass] 
\beq 
\tan\alpha \;=\; \frac{-(M_A^2+M_Z^2)\tan\beta}
  {M_Z^2+M_A^2\tan^2\beta-(1+\tan^2\beta)M_{h}}
\eeq
Once these parameters are fixed, the couplings of the Higgs particles to
gauge bosons and fermions will also be uniquely fixed; the Higgs--vector
boson couplings relevant to our analysis are given in Table 1a. \s

To fully describe the supersymmetric sector two parameters must be
introduced, in addition to $M_A, \tb, M_S, A_t,A_b$ and $\mu$. These are
the gaugino mass parameter $M_2$ [we will use the GUT relation which
fixes the bino mass in terms of the gaugino mass, $M_1= \frac{5}{3} 
\tan^2 \theta_W \, M_2$] and the common slepton mass $M_L$. The masses
of the four neutralinos $\chi_i^0$ [$i=1,..,4$] and the two charginos
$\chi_i^+$ [$i=1,2$] are then completely fixed by $\tb, \mu$ and $M_2$.
The masses of the left-- and right--handed selectron $\tilde{e}_L,
\tilde{e}_R$, and of the left--handed electronic sneutrino
$\tilde{\nu}_L$ will be fixed by $M_L$ [in practice these masses will
approximately be given by $m_{\tilde{e}_L} \simeq m_{\tilde{e}_R} \simeq
m_{\tilde{\nu}_L} =M_L$; the mixing between the two selectron states is
negligible]. \s

For the couplings of the Higgs bosons to neutralinos, charginos and
sleptons, one also will need the value of the mixing angle $\alpha$ 
which is fixed by $\tb$ and $M_A$. These couplings are given in Tables 
1b--1c; the couplings of the electrons to charginos/neutralinos and 
sleptons which will also be needed in the analysis are given in the 
Appendix. \s 

We are now in a position to discuss the numerical results. The cross
sections for the four processes $\ee {\ra} hh, HH, AA$ and $hH$ are
displayed  in Figs.3--6, as a function of the Higgs boson masses and for
two values of the c.m. energy $\sqrt{s}= 500 $ GeV and 1.5 TeV [except
for $hH$ production] and two
representative values of $\tb=1.5$ and 50. In all the figures we have
chosen the parameters $M_2=-\mu=150$ GeV, while the common slepton and
squark masses are taken to be $M_L=300$ GeV and $M_S=500$ GeV; the
parameter $A_t$ and $A_b$ are set to zero. Only the unpolarized cross
sections are discussed: as mentioned previously, they are simply increased by
a factor of 2(4) when the initial beams(s) are longitudinally polarized.
\s 

Fig.3 shows the cross section for the process $\ee {\ra} hh$. The solid
lines are for the full cross sections, while the dashed lines are for
the cross sections without the SUSY contributions [this will correspond 
to a two--Higgs doublet model with the MSSM constraints]. 
Let us first discuss
the case where the supersymmetric contributions are not included. 
For small $\tb$, the cross section is of
the same order as the SM cross section and does not strongly depend on 
$M_h$ especially at very high--energies. Although the $WWh/Zhh$ couplings
are suppressed by $\sin (\beta-\alpha)$ factors, the suppression is not
very strong and the $W/Z$ box contributions are not much smaller than in
the SM; the diagrams where $A/H^\pm$ are exchanged will give
compensating contributions since the $hAZ/hH^\pm W$ couplings are
proportional to the complementary factor $\cos(\beta- \alpha)$. As
in the SM case, the cross sections slightly decrease with increasing
energy. \s 

For large $\tb$ values, the factors $\sin /\cos(\beta-\alpha)$ vary
widely when $M_h$ is varied. For small $M_h$, the factor $\sin(\beta-\alpha)
\ra 0$, and the contribution of the diagrams with $A/H^\pm$
exchange dominates. The latter contribution decreases with increasing $M_h$
[i.e. with decreasing $\cos(\beta-\alpha)]$, until the decoupling limit is 
reached for $M_h \simeq 110$ GeV. In this case, the factor $\sin(\beta-\alpha)
\ra 1$ and the $W/Z$ boson loops are not suppressed anymore; one then 
obtains the SM cross section. \s


The contributions of the chargino/selectron and neutralino/sneutrino
loops lead to a destructive interference. At high--energies, the
supersymmetric boxes practically do not contribute; but at low energies,
and especially below the decoupling limit, the SUSY contributions can
be of the order of $\sim 10\%$. We have scanned the SUSY parameter
space, and the maximum contribution of the SUSY loops that we have found was
about $\sim -15\%$. In the decoupling limit, the SUSY contributions are,
at most, of the order of a  few percent. Because of the rather 
low production
rates, it will therefore be difficult to experimentally see this effect.
\s 

Fig.4 and 5 show the cross sections for the processes $\ee \ra HH$ and 
$\ee \ra AA$, respectively. Except for small $A/H$ masses where they are 
of the same order as in the SM, the cross sections are below the $0.1$ fb
level and the signals will be hard to detect, especially for $M_A, M_H
\gsim 350$ GeV [independently of the phase space suppression]. The SUSY
contributions are relatively even smaller than for the case of $\ee \ra
hh$. \s 

Finally, Fig.6 shows the cross section for the case $\ee \ra hH$
at $\sqrt{s} = 500 $ GeV. It
is one order of magnitude smaller than in the previous cases and
therefore negligibly small. This is due to the fact that the $W/Z$
and $A/H^\pm$ contributions are both suppressed since the combination
$\sin (\beta-\alpha) \times \cos( \beta-\alpha)$ appears in all these
contributions; since one of the factors is always small, this brings 
these contributions down to the level of the small SUSY--loop contributions. 

\subsection*{4. Summary}

We have analyzed the one--loop induced production of Higgs boson pairs at
future high--energy $\ee$ colliders  in the Standard Model and in
the Minimal Supersymmetric Standard Model. In the SM, the unpolarized cross
section is rather small, of the order $0.1$--$0.2$ fb. The longitudinal
polarization of both the $e^-$ and $e^+$ beams will increase the cross
section by a factor of 4. With integrated luminosities $\int {\cal L} \gsim 
100$ fb$^{-1}$ as expected to be the case for future high--energy
linear colliders, one could expect a few hundred events in the course of
a few years if longitudinal polarization is available. The final states 
are rather clean, 
giving a reasonable hope to isolate the signals experimentally. \s

In the MSSM, additional contributions to the processes $\ee \ra hh, HH,
Hh$ and $\ee \ra AA$ come from chargino/neutralino and slepton loops.
For $hh$ production, the contributions of the supersymmetric loops are
in general rather small, being of the order of a few percent; the cross
sections are therefore of the same order as in the SM. For the processes
involving heavy Higgs bosons, the cross sections are even smaller than
for $\ee \ra hh$, and the signals will be hard to be detected 
experimentally. 

\vspace*{1cm}

\nn {\bf Acknowledgements:} We thank W. Hollik for discussions and J. Rosiek 
for providing us with the Fortran code for the one--loop corrected 
MSSM Higgs boson masses. 

\vspace*{0.5cm} 

\setcounter{equation}{0}
\renewcommand{\theequation}{A\arabic{equation}}

\subsection*{APPENDIX: Cross sections in the MSSM}

In this Appendix, we will give the analytical expressions of the cross
sections for the pair production of the MSSM Higgs bosons. We start with
the production of two CP--even Higgs bosons, $\ee \ra \Phi_1 \Phi_2$
with $\Phi_1, \Phi_2 \equiv h,H$. Similarly to the SM case, the
differential cross section for the process can be written as: 
\begin{eqnarray} 
\frac{ \mbox{\rm d} \sigma}{\mbox{\rm d} \cos\theta} &=& 
\frac{1} {1+ \delta_{\Phi_1 \Phi_2} }
\frac{G_F^4\,M_W^8}{8\,\pi^5\, s} \,  \left[ \left( 1-
\frac{M_{\Phi_1}^2}{s} - \frac{M_{\Phi_2}^2}{s} \right)^2 -
4\frac{M_{\Phi_1}^2 M_{\Phi_2}^2 }{s^2} \right]^{1/2} \, (ut -
M_{\Phi_2}^2 M_{\Phi_1}^2 ) \non \\&& \mbox{} 
\left[ (1+\lambda_-)(1-\lambda_+)\left| \sum_{k=1}^{9} \,F^{+}_{k}\right|^2 
+ (1-\lambda_-)(1+\lambda_+)\left| \sum_{k=1}^{9} \,F^{-}_{k}\right|^2 
\right]
\end{eqnarray}
where $\lambda_-$ and $\lambda_+$ are the longitudinal polarizations of the
initial $e^-$ and $e^+$ beams, $\theta$ the scattering angle and $s,t,u$ 
the usual Mandelstam variables; $\delta_{\Phi_1 \Phi_2}=1$ in the case 
where we have two identical Higgs bosons in the final state, otherwise
$\delta_{\Phi_1 \Phi_2}=0$.  \s

For the class of diagrams consisting of $W/Z$ and $A/H^\pm$ loops [a two 
Higgs--doublet model], there are three contributing amplitudes 
$F^\pm_{1,2,3}$ given by 
\begin{eqnarray}
F_{1}^{+} &=&g_{\Phi_1VV} \,g_{\Phi_2VV} \frac{M_W^2}{2} \left[
\frac{(2s_W^2-1)^2}{2c_W^6}F_{A}(M_Z,0,M_Z,M_Z,s,t) 
+F_{A}(M_W,0,M_W,M_W,s,t) 
\right]
\non \\
F_{1}^{-} &=& g_{\Phi_1VV}\, g_{\Phi_2VV}\,\frac{{M_Z^2}\,s_W^4}{c_W^4}
\;F_{A}(M_Z,0,M_Z,M_Z,s,t) 
\non \\
F_{2}^{+} &=& g_{\Phi_1VV}\, g_{\Phi_2VV}\,\left[
\frac{(2s_W^2-1)^2}{16\,c_W^4}\;F_{B}(M_Z,0,M_Z,M_Z,s,t) 
+\frac{1}{8}\;F_{B}(M_W,0,M_W,M_W,s,t) 
\right]
\non \\
F_{2}^{-} &=& g_{\Phi_1VV}\, g_{\Phi_2VV}\,\frac{s_W^4}{4\,c_W^4}
\;F_{B}(M_Z,0,M_Z,M_Z,s,t) 
\non \\
F_{3}^{+} &=& g_{\Phi_1AV}\, g_{\Phi_2AV}\;
\frac{(2s_W^2-1)^2}{16\,c_W^4}F_{B}(M_Z,0,M_Z,M_A,s,t) 
\non \\ && \mbox{} 
+g_{\Phi_1 H^{\pm} W^{\pm}} \;g_{\Phi_2 H^{\pm} W^{\pm}}\; 
 \frac{1}{8}F_{B}(M_W,0,M_W,M_{H^\pm},s,t) 
\non \\
F_{3}^{-} &=& g_{\Phi_1AV}\, g_{\Phi_2AV}\,\frac{s_W^4}{4\,c_W^4}
\;F_{B}(M_Z,0,M_Z,M_A,s,t) 
\end{eqnarray}
where the couplings $g_{\Phi VV}, g_{\Phi AZ}$ and $g_{\Phi H^\pm W}$
are given in Table 1a. In terms of the $D_{ijk}$ four--point
Passarino--Veltman functions \cite{R13}, the form factors $F_{A,B} \equiv
F_{A,B}(m_1,m_2,m_3,m_4,$ $s,t)$ [the analogous of $F_1$ and $F_2$ of
eqs.(7) and (8)] are given by 
\begin{eqnarray}
{F_{A}} &=& {D_{13}}\;+\;\{u\leftrightarrow t \ , \ 
M_{\Phi_1} \leftrightarrow M_{\Phi_2}\}
\\[3mm]
{F_{B}} &=& \Bigg[ 2(M_{\Phi_2}^2 -u)\,{D_0}  
+ 2(M_{\Phi_2}^2 -u)\,{D_{11}}+ 2(M_{\Phi_2}^2 -t)\,{D_{12}} 
+ 2\,{t}\,{D_{13}} 
\\ && 
+  \frac {3\,M_{\Phi_1}^2 + t }{2}\,{D_{23}} + 2\,{s}\,{D_{24}} 
+ 2\,(\,t - M_{\Phi_1}^2 \,)\,{D_{25}} 
+  \frac{s + 4\,(u-M_{\Phi_1}^2)}{2}\,{D_{26}} + 8\,{D_{27}} 
\non \\ &&
+ {\frac {M_{\Phi_1}^2}{2}}\,{D_{33}} 
+  \frac{ t - \,M_{\Phi_1}^2 }{2} \,{D_{37}} 
+  \frac{ u - \,M_{\Phi_1}^2 }{2} \,{D_{39}} 
+ {\frac {s}{2}}\,{D_{310}} + 3\,{D_{313}} 
\Bigg] \non \\ &&
\;+\;\{u\leftrightarrow t \ , \
M_{\Phi_1} \leftrightarrow M_{\Phi_2}\} \non 
\end{eqnarray}
with
\[
D_{ijk} \;\equiv \; D_{ijk}(m_1,m_2,m_3,m_4,0,0,M_{\Phi_1},M_{\Phi_2},s,t)
\]

For the class of diagrams consisting of sneutrino/chargino loops,
the contributing amplitudes $F^\pm_{4,5,6}$ are given by 
\begin{eqnarray}
F_{4}^{+} &=& -\sum_{i=1}^{2}\frac{M_Z^2}{8\,c_W^2}\,
 g_{e\chi^+_i\tilde{\nu}}\;g_{e\chi^+_i\tilde{\nu}}\;
  g_{\Phi_1\tilde{\nu}\tilde{\nu}}\;g_{\Phi_2\tilde{\nu}\tilde{\nu}} \; 
   F_{A}(m_{\tilde{\nu}},-m_{\chi_i^+},m_{\tilde{\nu}},m_{\tilde{\nu}},s,t)
\non \\[2mm]
F_{5}^{+} &=& \sum_{i,j=1}^{2}\,
 \frac{M_Z}{2\,c_W}\, g_{e\chi^+_i\tilde{\nu}}\;g_{e\chi^+_j\tilde{\nu}}
\non\\&&
\sum_{a=L,R} \Big[ 
  g_{\Phi_1\tilde{\nu}\tilde{\nu}}\;g_{\Phi_2\chi^{+}_i\chi^{-}_j}^a\;
   F^{+a}_{C}(-m_{\chi_i^+},m_{\tilde{\nu}},m_{\tilde{\nu}},-m_{\chi_j^+},s,t,
   M_{\Phi_1},M_{\Phi_2}) 
\non\\&&\hspace{.8cm}
+ g_{\Phi_2\tilde{\nu}\tilde{\nu}}\;g_{\Phi_1\chi^{+}_i\chi^{-}_j}^a\;
  F^{+a}_{C}
(-m_{\chi_i^+},m_{\tilde{\nu}},m_{\tilde{\nu}},-m_{\chi_j^+},s,u,
   M_{\Phi_2},M_{\Phi_1}) 
\Big]
\non \\[2mm]
F_{6}^{+} &=& - \sum_{i,j,k=1}^{2}\;
 g_{e\chi^+_i\tilde{\nu}}\;g_{e\chi^+_k\tilde{\nu}}
\non\\&&
 \,\sum_{a,b=L,R} \Big[ 
  g_{\Phi_1\chi^{+}_i\chi^{-}_j}^a\;g_{\Phi_2\chi^{+}_j\chi^{-}_k}^{b}\;
  F^{+ab}_{D}(-m_{\chi_i^+},m_{\tilde{\nu}},-m_{\chi_k^+},-m_{\chi_j^+},s,t,
   M_{\Phi_1},M_{\Phi_2}) 
\non\\&&\hspace{1cm}
+ g_{\Phi_2\chi^{+}_i\chi^{-}_j}^a\;g_{\Phi_1\chi^{+}_j\chi^{-}_k}^{b}\;
 F^{+ab}_{D}
  (-m_{\chi_i^+},m_{\tilde{\nu}},-m_{\chi_k^+},-m_{\chi_j^+},s,u,
   M_{\Phi_2},M_{\Phi_1}) 
\Big] 
\end{eqnarray}
and
\begin{eqnarray}
F_{4,5,6}^{-} = 0 
\end{eqnarray}
$F_A$ is given in eq.(A3), and the new form factors $F^{\pm a}_{C} \equiv 
F^{\pm a} _{C}(m_1,m_2,m_3,m_4,s,t,M_{\Phi_1},M_{\Phi_2})$ and
$F^{\pm ab}_{D} \equiv 
F^{\pm ab}_{D}(m_1,m_2,m_3,m_4,s,t,M_{\Phi_1},M_{\Phi_2})$,
with $a,b=L,R$, are given by 
\begin{eqnarray}
F^{-R}_{C} \;=\;\, F^{+L}_{C}\; 
&=& {\frac {1}{2}}\,\big(\,{m_1}\,{D_0}+{m_1}\,{D_{12}} \big) \non\\
F^{-L}_{C} \;=\;\, F^{+R}_{C}\;
&=& \frac {1}{2} \,{m_4}\,{D_{12}} \non\\
F^{-RR}_{D} \;=\; F^{+LL}_{D} 
&=& -{\frac {1}{2}}\, {m_1}\,{m_4} \,{D_{13}} \non\\
F^{-RL}_{D} \;=\; F^{+LR}_{D}
&=& { \frac {1}{2}}\,\Big[ 
 {s}\,({D_{12}} +{D_{13}} ) + ({s} -M_{\Phi_2}^{2})\,{D_{23}} 
+ {s} \,( {D_{24}} - {D_{25}} - {D_{26}}) + 2\,{D_{27}} 
\non\\ && \mbox{}
- M_{\Phi_1}^{2}\,{D_{33}} - ({t} - M_{\Phi_1}^{2})\,{D_{37}} 
-({u} -M_{\Phi_1}^{2})\,{D_{39}} -{s}\,{D_{310}} -6\,{D_{313}} \Big] \non\\
F^{-LR}_{D} \;=\;F^{+RL}_{D} 
&=& -{\frac {1}{2}}\,{m_1}\,{m_3}\,
    (\, {D_0} + {D_{13}} \,)  \non\\ 
F^{-LL}_{D} \;=\; F^{+RR}_{D}
&=& -{\frac {1}{2}}\,{m_3}\,{m_4}\,{D_{13}}
\end{eqnarray}

\smallskip

Finally, for the class of diagrams consisting of selectron/neutralino 
loops, the amplitudes $F^\pm_{7,8,9}$ are given by 
\begin{eqnarray}
F_{7}^{+} &=& \sum_{i=1}^{4}\,\frac{M_Z^2}{8\,c_W^2}\,
 g_{e\chi^0_i\tilde{e}_R}\;g_{e\chi^0_i\tilde{e}_R}\;
 g_{\Phi_1\tilde{e}_R\tilde{e}_R} \;g_{\Phi_2\tilde{e}_R\tilde{e}_R} \;
  F_{A}(m_{\tilde{e}_R},m_{\chi_{i}^0},m_{\tilde{e}_R},m_{\tilde{e}_R},s,t) 
\non \\[3mm]
F_{8}^{+} &=& \sum_{i,j=1}^{4}\,\frac{M_Z}{4\,c_W}\,
 g_{e\chi^0_i\tilde{e}_R}\;g_{e\chi^0_j\tilde{e}_R}\;
\non  \\&&\mbox{}\hspace{1cm}
\sum_{a=L,R} \Big[
  g_{\Phi_1\tilde{e}_R\tilde{e}_R}\;g_{\Phi_2\chi^{0}_i\chi^{0}_j}^a\;
 F^{+a}_{C}(m_{\chi_{i}^0},m_{\tilde{e}_R},m_{\tilde{e}_R},m_{\chi_{j}^0},s,t,
     M_{\Phi_1},M_{\Phi_2})
\non  \\&&\mbox{}\hspace{1.7cm}
 + g_{\Phi_2\tilde{e}_R\tilde{e}_R}\;g_{\Phi_1\chi^{0}_i\chi^{0}_j}^a\;
 F^{+a}_{C}(m_{\chi_{i}^0},m_{\tilde{e}_R},m_{\tilde{e}_R},m_{\chi_{j}^0},s,u,
     M_{\Phi_2},M_{\Phi_1})
\Big]
\non \\[3mm]
F_{9}^{+} &=& \sum_{i,j,k=1}^{4}\,\frac{1}{2}\,
 g_{e\chi^0_i\tilde{e}_R}\;g_{e\chi^0_k\tilde{e}_R}\;
\non \\&&\mbox{}\hspace{1cm}
\sum_{a,b=L,R}\,\Big[ 
   g_{\Phi_1\chi^{0}_i\chi^{0}_j}^a\;g_{\Phi_2\chi^{0}_j\chi^{0}_k}^b\;
 F^{+ab}_{D}(m_{\chi_{i}^0},m_{\tilde{e}_R},m_{\chi_{k}^0},
m_{\chi_{j}^0},s,t,
      M_{\Phi_1},M_{\Phi_2})
\non \\&&\mbox{}\hspace{2cm}
  + g_{\Phi_2\chi^{0}_i\chi^{0}_j}^a\;g_{\Phi_1\chi^{0}_j\chi^{0}_k}^b\;
 F^{+ab}_{D}
  (m_{\chi_{i}^0},m_{\tilde{e}_R},m_{\chi_{k}^0},m_{\chi_{j}^0},s,u,
      M_{\Phi_2},M_{\Phi_1}) \Big]
\end{eqnarray}
and
\begin{eqnarray}
F_{7,8,9}^{-} = F_{7,8,9}^{+} (\tilde{e}_R \to \tilde{e}_L \ , 
\ F^{+ab}_{C,D} \ra F^{-ab}_{C,D} ) 
\end{eqnarray}
where the form factors $F_{A,B,C,D}$ have been given previously. The
normalized couplings of the neutral Higgs bosons to charginos,
neutralinos and sleptons are displayed in Tab.~1b and 1c. The only
remaining couplings which have to be defined, are the
electron--chargino--sneutrino and electron--neutralino--selectron
couplings, which, normalized to $g_{W}=\left[\sqrt{2}G_F\right]^{ 1/2}$
$M_W$, are given by 
\beq
g_{e \chi_i^+ \tilde{\nu}} = V_{i1} \ \ , \ \ 
g_{e \chi_i^0 \tilde{e}_R} = -2\frac{s_W}{c_W} N_{i2} \ \ , \ \ 
g_{e \chi_i^0 \tilde{e}_L} = -N_{i2} -\frac{s_W}{c_W} N_{i1} 
\eeq
where the matrices $V$ and $N$ can be found in Ref.\cite{R21}.  

\bigskip

For the pair production of the CP--odd Higgs boson in the MSSM, $\ee \ra
AA$, the differential cross section takes the form
\begin{eqnarray}
\frac{ \mbox{\rm d} \sigma}{\mbox{\rm d} \cos\theta} &=&
\frac{G_F^4\,M_W^8}{16\,s\,\pi^5} \sqrt{1-4\frac{M_A^2}{s}} \; (ut - M_A^4)
\non \\&& \mbox{}
\left[ (1+\lambda_-)(1-\lambda_+)\left| \sum_{k=10}^{12} \,F^{+}_{k}\right|^2 
+ (1-\lambda_-)(1+\lambda_+)\left| \sum_{k=10}^{12} \,F^{-}_{k}\right|^2 
\right]
\end{eqnarray}
where we have used the same notation as previously. The three contributing 
amplitudes $F^\pm_{10,11,12}$ are given by 
\begin{eqnarray}
F_{10}^{+} &=& 
\frac{(2s_W^2-1)^2}{16\,c_W^4}\;\Big[
g^2_{hAV} \, F_{B}(M_Z,0,M_Z,M_h,s,t) 
\non \\&&\mbox{} 
+ g^2_{HAV} \,F_{B}(M_Z,0,M_Z,M_H,s,t)\Big]
+\frac{1}{8} g^2_{H^\pm AW} 
\;F_{B}(M_W,0,M_W,M_{H^\pm},s,t) 
\non \\[3mm]
F_{10}^{-} &=& \frac{s_W^4}{4\,c_W^4} \left[
g^2_{hAV} F_{B}(M_Z,0,M_Z,M_h,s,t) 
+g^2_{HAV} F_{B}(M_Z,0,M_Z,M_H,s,t) \right] \non \\[3mm]
F_{11}^{+} &=& \sum_{i,j,k=1}^{4}\,\frac{1}{2}\,
 g_{e\chi^0_i\tilde{e}_R}\;g_{e\chi^0_k\tilde{e}_R}\;
\non \\&&\mbox{}\hspace{1cm}
 \sum_{a,b=L,R}\,\Big[
 g_{A\chi^{0}_i\chi^{0}_j}^a\;g_{A\chi^{0}_j\chi^{0}_k}^b\;
 F^{+ab}_{D}
 (m_{\chi_{i}^0},m_{\tilde{e}_R},m_{\chi_{k}^0},m_{\chi_{j}^0},s,t,M_{A},M_{A})
\non \\&&\mbox{}\hspace{2cm}
+ g_{A\chi^{0}_i\chi^{0}_j}^a\;g_{A\chi^{0}_j\chi^{0}_k}^b\;
 F^{+ab}_{D}
  (m_{\chi_{i}^0},m_{\tilde{e}_R},m_{\chi_{k}^0},m_{\chi_{j}^0},
  s,u,M_{A},M_{A})\Big]
\non \\[3mm]
F_{11}^{-} &=& \sum_{i,j,k=1}^{4}\,\frac{1}{2}\,
 g_{e\chi^0_i\tilde{e}_L}\;g_{e\chi^0_k\tilde{e}_L}\;
\non \\&&\mbox{}\hspace{1cm}
 \sum_{a,b=L,R}\,\Big[
 g_{A\chi^{0}_i\chi^{0}_j}^a\;g_{A\chi^{0}_j\chi^{0}_k}^b\;
 F^{-ab}_{D}
 (m_{\chi_{i}^0},m_{\tilde{e}_L},m_{\chi_{k}^0},m_{\chi_{j}^0},s,t,M_{A},M_{A})
\non \\&&\mbox{}\hspace{2cm}
+ g_{A\chi^{0}_i\chi^{0}_j}^a\;g_{A\chi^{0}_j\chi^{0}_k}^b\;
 F^{-ab}_{D}(m_{\chi_{i}^0},m_{\tilde{e}_L},m_{\chi_{k}^0},m_{\chi_{j}^0},
    s,u,M_{A},M_{A})\Big]
\non \\[3mm]
F_{12}^{+} &=& - \sum_{i,j,k=1}^{2}\, 
g_{e\chi^+_i\tilde{\nu}}\;g_{e\chi^+_k\tilde{\nu}}
\non \\&&\mbox{}\hspace{1cm}
 \sum_{a,b=L,R}\,\Big[
 g_{A\chi^{+}_i\chi^{-}_j}^a\;g_{A\chi^{+}_j\chi^{-}_k}^b\;
  F^{+ab}_{D}(-m_{\chi_i^+},m_{\tilde{\nu}},-m_{\chi_k^+},-m_{\chi_j^+},s,t,
     M_{A},M_{A})
\non \\&&\mbox{}\hspace{2cm}
+ g_{A\chi^{+}_i\chi^{-}_j}^a\;g_{A\chi^{+}_j\chi^{-}_k}^b\;
  F^{+ab}_{D}(-m_{\chi_i^+},m_{\tilde{\nu}},-m_{\chi_k^+},-m_{\chi_j^+},s,u,
     M_{A},M_{A}) \Big]
\non \\[3mm]
F_{12}^{-} &=& 0
\end{eqnarray}
with $F_{B}$ and $F_{D}$ as given previously; 
the couplings of the pseudoscalar Higgs bosons to 
the neutralino and chargino states are again given in Table 1c. 

\newpage

%
\def\npb#1#2#3{{\rm Nucl. Phys. }{\rm B #1} (#2) #3}
\def\plb#1#2#3{{\rm Phys. Lett. }{\rm #1 B} (#2) #3}
\def\prd#1#2#3{{\rm Phys. Rev. }{\rm D #1} (#2) #3}
\def\prl#1#2#3{{\rm Phys. Rev. Lett. }{\rm #1} (#2) #3}
\def\prc#1#2#3{{\rm Phys. Rep. }{\rm C #1} (#2) #3}
\def\pr#1#2#3{{\rm Phys. Rep. }{\rm #1} (#2) #3}
\def\zpc#1#2#3{{\rm Z. Phys. }{\rm C #1} (#2) #3}
\def\nca#1#2#3{{\it Nouvo~Cim.~}{\bf #1A} (#2) #3}
%


\newpage

\section*{ Table 1:}

\begin{center}
\begin{tabular}{|c||c||c|c|} \hline
& & & \\ 
$\hspace{1cm} \Phi \hspace{1cm}$ &  $g_{\Phi VV} $ &
$ g_{\Phi AZ} $ & $g_{\Phi H^\pm W^\pm} $ \\[0.3cm] \hline \hline & & & \\
$h$ &  $ \ \sin(\beta-\alpha) \ $ & \ $ \; \ \cos (\beta-\alpha)  \; 
$ \ & \ $ \;
\mp \cos (\beta-\alpha)  \; $ \\
$H$ & \ $ \; \cos (\beta-\alpha) \; $ &
\ $ \; - \sin (\beta-\alpha) \; $ \ & \ $ \;
\pm \sin (\beta-\alpha)  \; $ \\
$A$ & $0$ & \ $ \;   0 \; $ \ & \ $ \; 1  \; $ \\[0.3cm] \hline
\end{tabular}
\end{center}

\vspace*{3mm}

\nn {\small {\bf Tab.~1a}: The Higgs--vector boson couplings $g_{\Phi VV}$
[normalized the SM Higgs coupling $g_{H^0VV}=2 \left[
\sqrt{2}G_F\right]^{1/2} M_V^2$], and the Higgs--Higgs--vector boson
couplings [normalized to $g_W=( \sqrt{2} G_F)^{1/2}$ $M_W$ and
$g_Z=(\sqrt{2}G_F)^{1/2}M_Z$ for the charged/neutral weak couplings];
the latter come with the sum of the Higgs momenta entering and leaving
the vertices.} 

\bigskip

\begin{center}
\begin{tabular}{|c||c|c|c|} \hline
& & & \\ 
$\hspace{1cm} \tilde{l}_i \tilde{l}_j \hspace{1cm}$ 
&  $g_{h \tilde{l}_i \tilde{l}_j} $ 
&  $g_{H \tilde{l}_i \tilde{l}_j} $ 
&  $\ \ g_{A \tilde{l}_i \tilde{l}_j} \ \ $ \\[0.3cm] \hline \hline & & & \\
$\tilde{e}_L \tilde{e}_L $ 
& $ \ (2s_W^2-1) \sin(\beta+\alpha) \ $ 
& $ \ -(2s_W^2-1) \cos(\beta+\alpha) \ $ 
& $ 0$ \\
$\tilde{e}_R \tilde{e}_R $ 
& $ \ 2s_W^2 \sin(\beta+\alpha) \ $ 
& $ \ -2s_W^2  \cos(\beta+\alpha) \ $ 
& $ 0$ \\
$\tilde{\nu}_L \tilde{\nu}_L $ 
& $ \ \sin(\beta+\alpha) \ $ 
& $ \ - \cos(\beta+\alpha) \ $ 
& $ 0$ \\[0.3cm] \hline
\end{tabular}
\end{center}

\vspace*{3mm}

\nn {\small {\bf Tab.~1b}: The couplings of the neutral Higgs bosons to left-- 
and right--handed sleptons, normalized to $g_{W}'= \left[\sqrt{2}G_F\right]^{
1/2} M_W^2$.} 

\bigskip

\begin{center}
\begin{tabular}{|c||c|c|c|} \hline
& & & \\ 
$\hspace{.2cm} g^{L,R} \ / \ \Phi \hspace{.2cm}$ 
&  $h$  &  $H$  &  $A$  \\[0.3cm] \hline \hline & & & \\
$g^L_{\Phi \chi^+_i \chi_j^-}$ & 
$Q^*_{ji}\sin\alpha - S^*_{ji}\cos\alpha $ &
$-Q^*_{ji}\cos\alpha - S^*_{ji}\sin\alpha $ &
$-Q^*_{ji}\sin\beta - S^*_{ji}\cos\beta$  \\
$g^R_{\Phi \chi^+_i \chi_j^-}$ & 
$Q_{ij}\sin\alpha - S_{ij}\cos\alpha $ &
$-Q_{ij}\cos\alpha - S_{ij}\sin\alpha $ &
$Q_{ij}\sin\beta + S_{ij}\cos\beta$  
\\ \hline
$g^L_{\Phi \chi^0_i \chi_j^0}$ & 
$Q^{"*}_{ji}\sin\alpha + S^{"*}_{ji}\cos\alpha $ &
$-Q^{"*}_{ji}\cos\alpha + S^{"*}_{ji}\sin\alpha $ &
$-Q^{"*}_{ji}\sin\beta + S^{"*}_{ji}\cos\beta$  \\
$g^R_{\Phi \chi^0_i \chi_j^0}$ & 
$Q^"_{ij}\sin\alpha + S^"_{ij}\cos\alpha $ &
$-Q^"_{ij}\cos\alpha + S^"_{ij}\sin\alpha $ &
$Q^"_{ij}\sin\beta - S^"_{ij}\cos\beta$  \\[0.3cm] \hline
\end{tabular}
\end{center}

\vspace*{3mm}

\nn {\small {\bf Tab.~1c}: The couplings of the neutral Higgs bosons to
charginos and neutralinos, normalized to $g_{W}=
\left[\sqrt{2}G_F\right]^{ 1/2} M_W$; the matrix elements $Q_{ij}/
S_{ij}$ and $Q^{"}_{ij}/S^{"}_{ij}$ can be found in Ref.~\cite{R21}.} 

\newpage

\nn {\Large \bf Figure Captions}

\begin{itemize}

\item[{\bf Fig.~1:~}]
Feynman diagrams contributing to the Higgs boson pair production process in
$\ee$ collisions: (a) SM Higgs production, CP--even (b) and CP--odd (c) Higgs 
boson production in the MSSM. 

\item[{\bf Fig.~2:}] a) The cross sections for the Higgs pair
production in the SM, $\ee \ra H^0 H^0$, as a function of the Higgs mass
for two center--of--mass energies, $\sqrt{s}=500$ GeV [dashed lines] and
$\sqrt{s}=1.5$ TeV [solid lines]. The lower curves correspond to the
unpolarized cross sections, the middle curves to the cross sections where
the electron beam is longitudinally polarized and the upper curves to
the cross sections where both the electron and positron beams are
longitudinally polarized. (b) The angular distribution as a function
of the scattering angle in the unpolarized case, at $\sqrt{s}=500$ GeV 
and for $M_H=150$ GeV. 

\item[{\bf Fig.~3:}] The cross sections for the pair production 
of the lightest Higgs boson in the MSSM, $\ee \ra hh$, as a function of 
$M_h$ for two center--of--mass energies, $\sqrt{s}=500$ GeV and
$\sqrt{s}=1.5$ TeV and for two values of $\tb=1.5$ and $50$.
The solid curves correspond to the full cross sections, while the dashed 
curves correspond to the cross sections without the SUSY contributions. 

\item[{\bf Fig.~4:}] The cross sections for the process $\ee \ra HH$, as a 
function of $M_H$ for two center--of--mass energies, $\sqrt{s}=500$ GeV and
$\sqrt{s}=1.5$ TeV and for two values of $\tb=1.5$ and $50$.
The solid curves correspond to the full cross sections, while the dashed 
curves correspond to the cross sections without the SUSY contributions. 

\item[{\bf Fig.~5:}] The cross sections for the process $\ee \ra AA$, as a 
function of $M_A$ for two center--of--mass energies, $\sqrt{s}=500$ GeV and
$\sqrt{s}=1.5$ TeV and for two values of $\tb=1.5$ and $50$.
The solid curves correspond to the full cross sections, while the dashed 
curves correspond to the cross sections without the SUSY contributions. 

\item[{\bf Fig.~6:}] The cross sections for the process $\ee \ra Hh$, as
a function of $M_H$ for $\sqrt{s}=500$ GeV and for two values of
$\tb=1.5$ and $50$. The solid curves correspond to the full
cross sections, while the dashed curves correspond to the cross sections
without the SUSY contributions. 

\end{itemize}

\newpage
\begin{figure}[ht]
\begin{center}
\mbox{
\psfig
 {figure=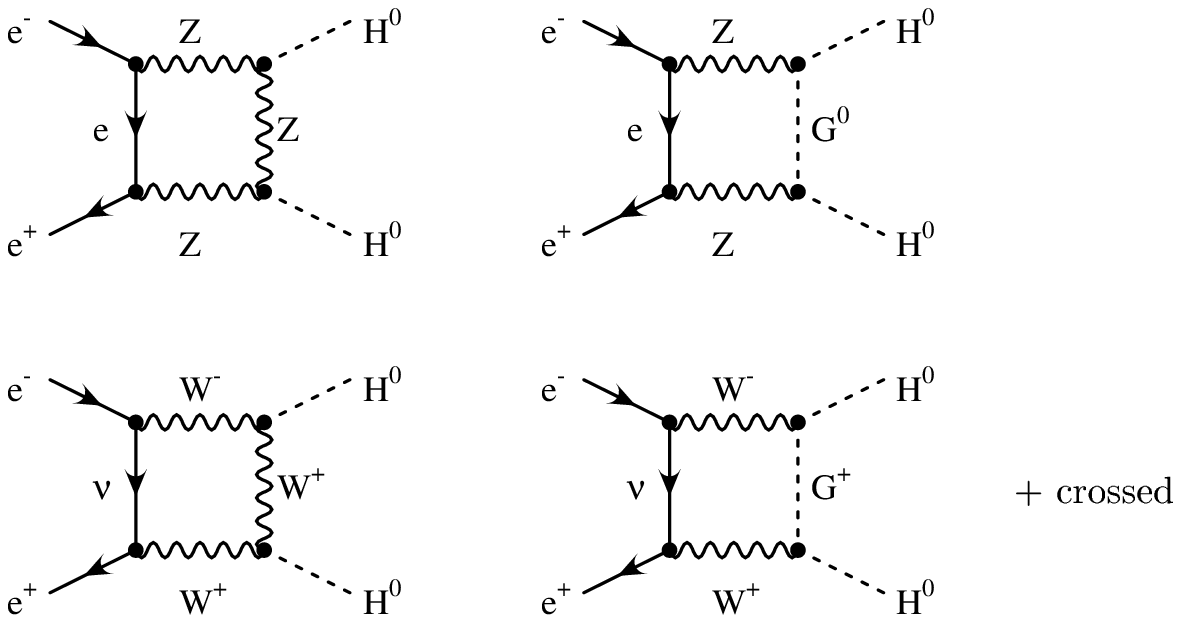,width=13cm,bbllx=90pt,bblly=260pt,bburx=450pt,bbury=460pt}
}
\end{center}                                
\centerline{\bf Fig. 1a}
\end{figure}                                
\begin{figure}[ht]
\begin{center}
\mbox{
\psfig
 {figure=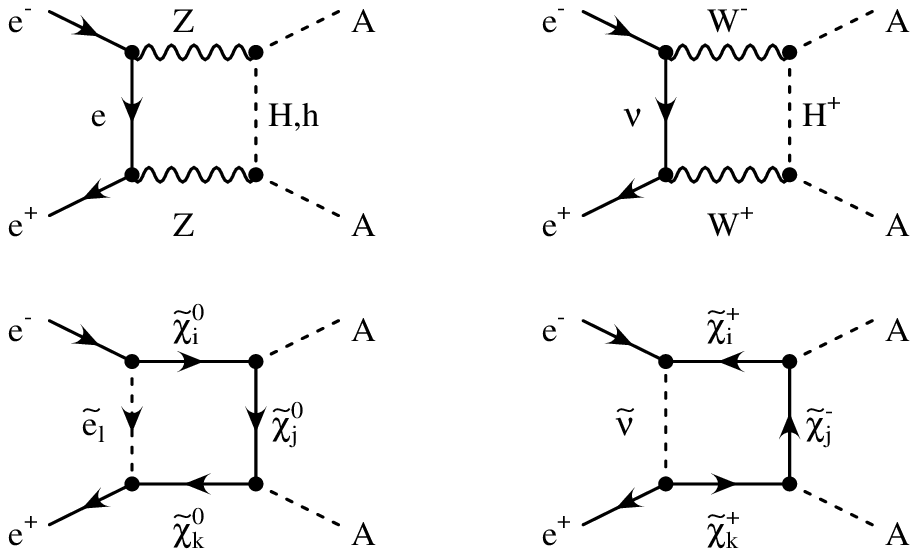,width=10cm,bbllx=130pt,bblly=260pt,bburx=400pt,bbury=460pt}
}
\end{center}                                
\centerline{\bf  Fig. 1c
}\label{fig_aa}
\end{figure}                                
\begin{figure}[ht]
\begin{center}
\mbox{
\psfig
 {figure=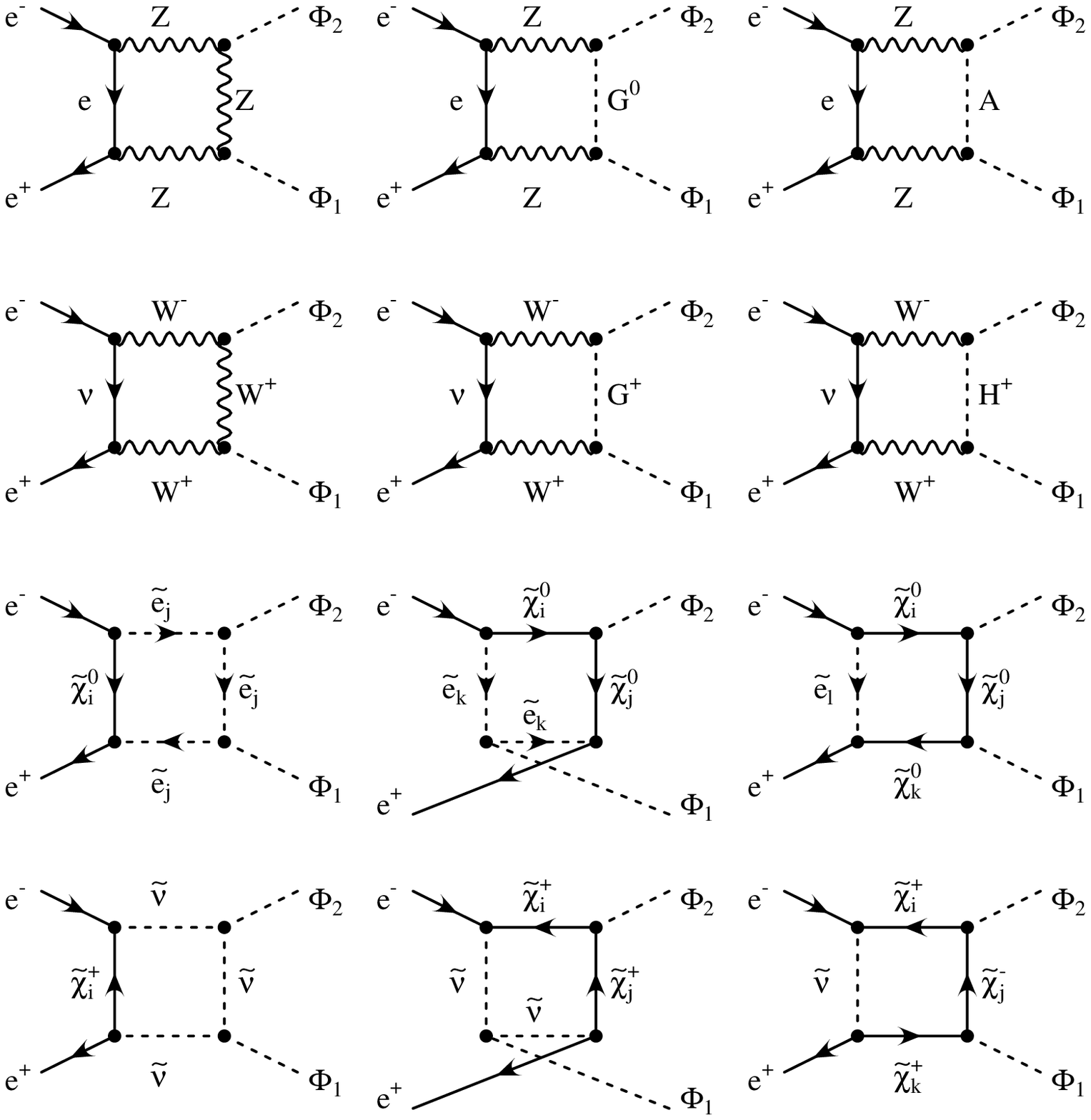,width=15cm,bbllx=60pt,bblly=120pt,bburx=530pt,bbury=600pt}
}
\end{center}                                
\centerline{\bf  Fig. 1b
}\label{fig_hh}
\end{figure}                                
\begin{figure}[ht]
\begin{center}
\mbox{
\psfig
 {figure=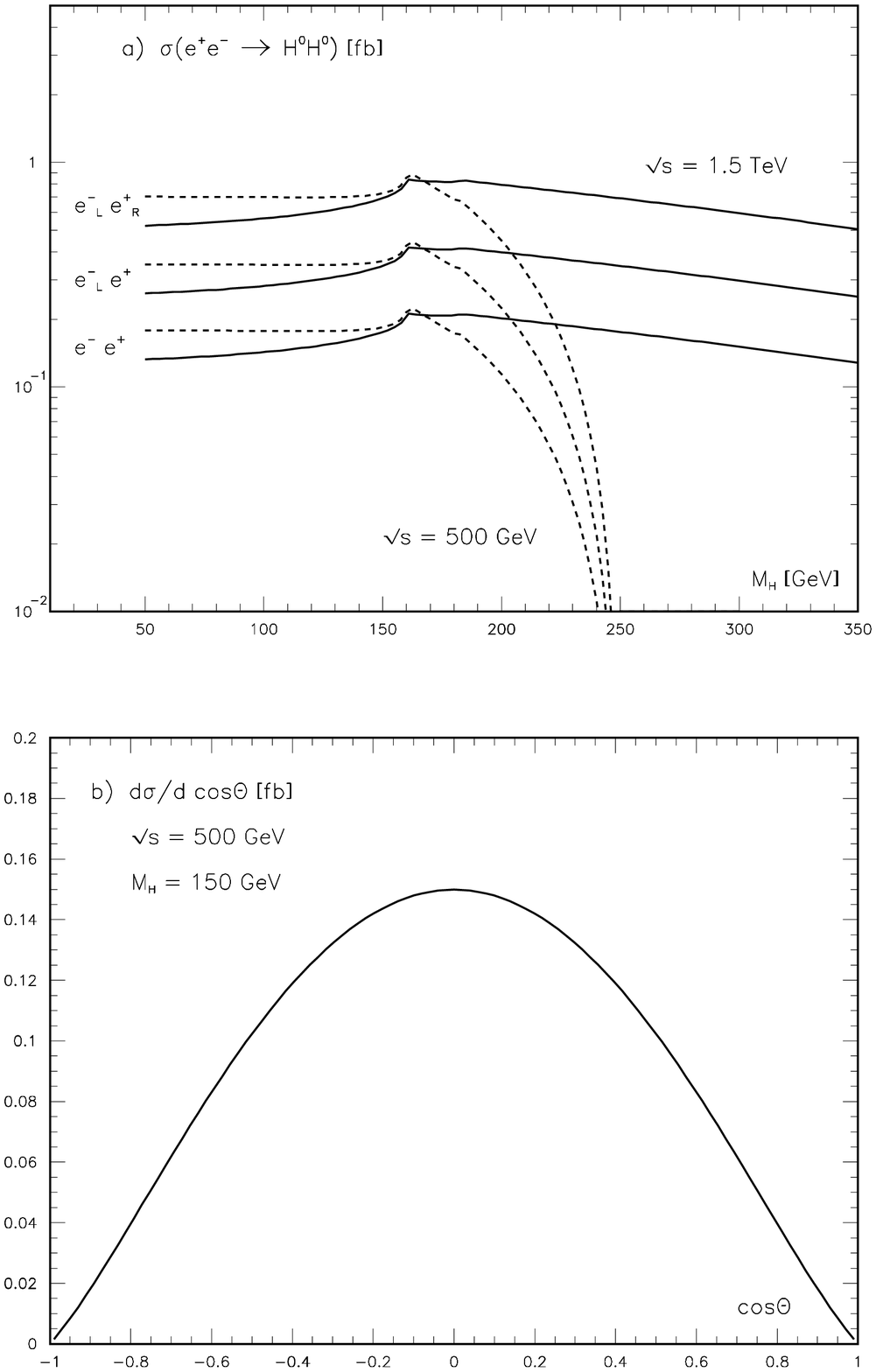,width=15cm,bbllx=30pt,bblly=50pt,bburx=550pt,bbury=760pt}
}
\end{center}                                
\centerline{\bf  Fig. 2
}\label{fig_sm}
\end{figure}                                
\begin{figure}[ht]
\begin{center}
\mbox{
\psfig
 {figure=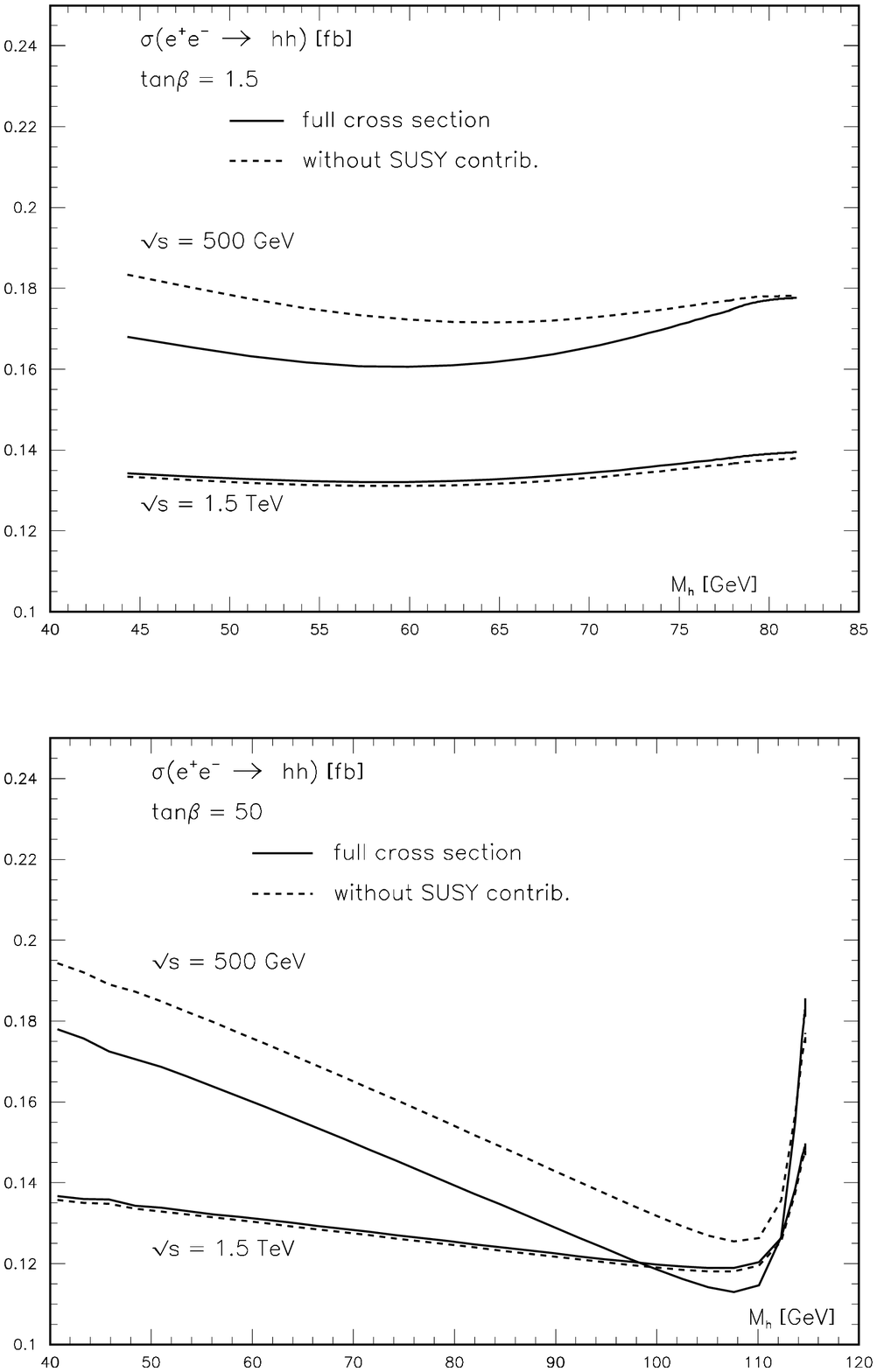,width=15cm,bbllx=30pt,bblly=50pt,bburx=550pt,bbury=760pt}
}
\end{center}                                
\centerline{\bf  Fig. 3
}\label{fig_s1}
\end{figure}                                
\begin{figure}[ht]
\begin{center}
\mbox{
\psfig
 {figure=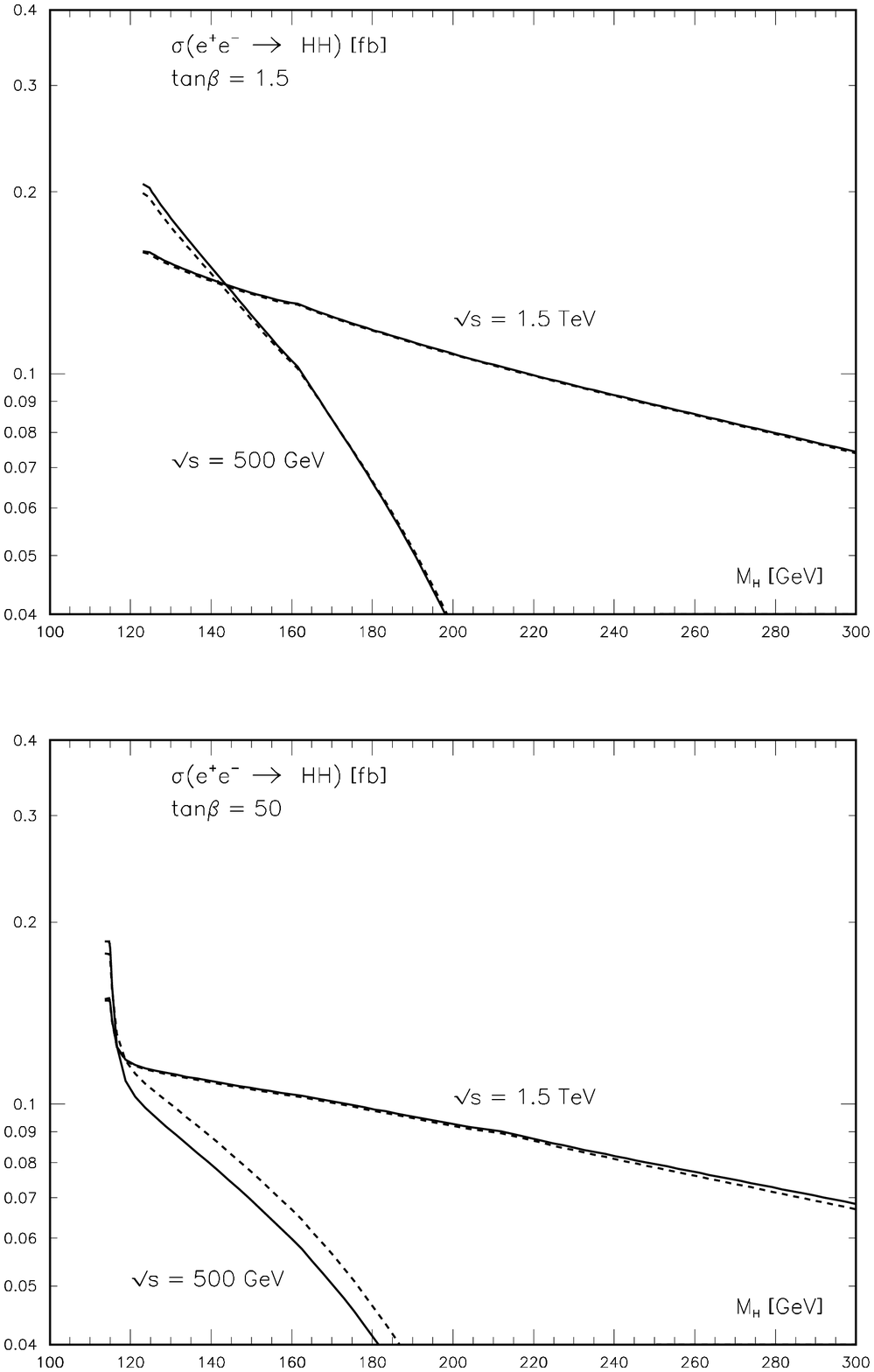,width=15cm,bbllx=30pt,bblly=50pt,bburx=550pt,bbury=760pt}
}
\end{center}                                
\centerline{\bf  Fig. 4
}\label{fig_s2}
\end{figure}                                
\begin{figure}[ht]
\begin{center}
\mbox{
\psfig
 {figure=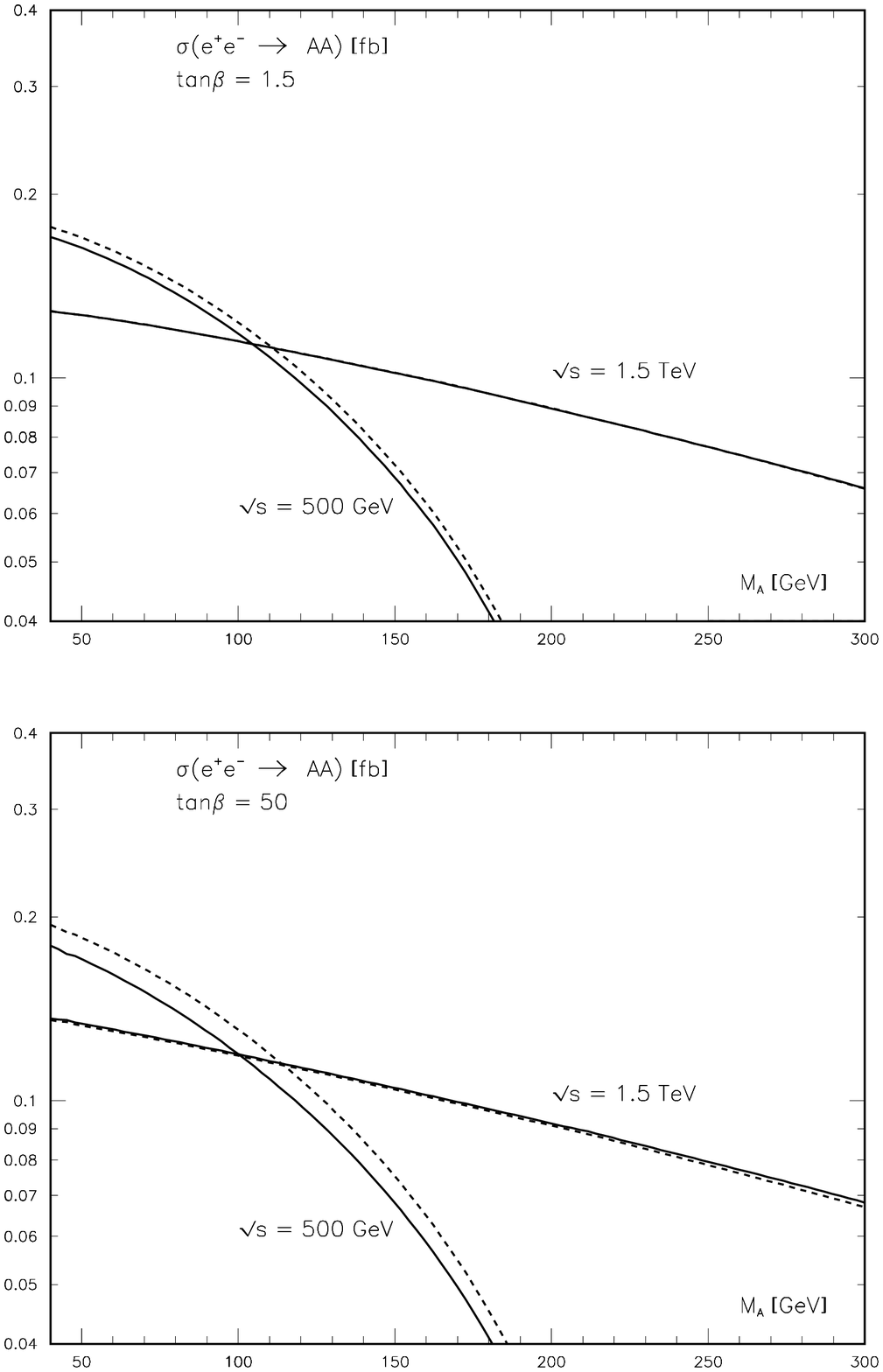,width=15cm,bbllx=30pt,bblly=50pt,bburx=550pt,bbury=760pt}
}
\end{center}                                
\centerline{\bf  Fig. 5
}\label{fig_s3}
\end{figure}                                
\begin{figure}[ht]
\begin{center}
\mbox{
\psfig
 {figure=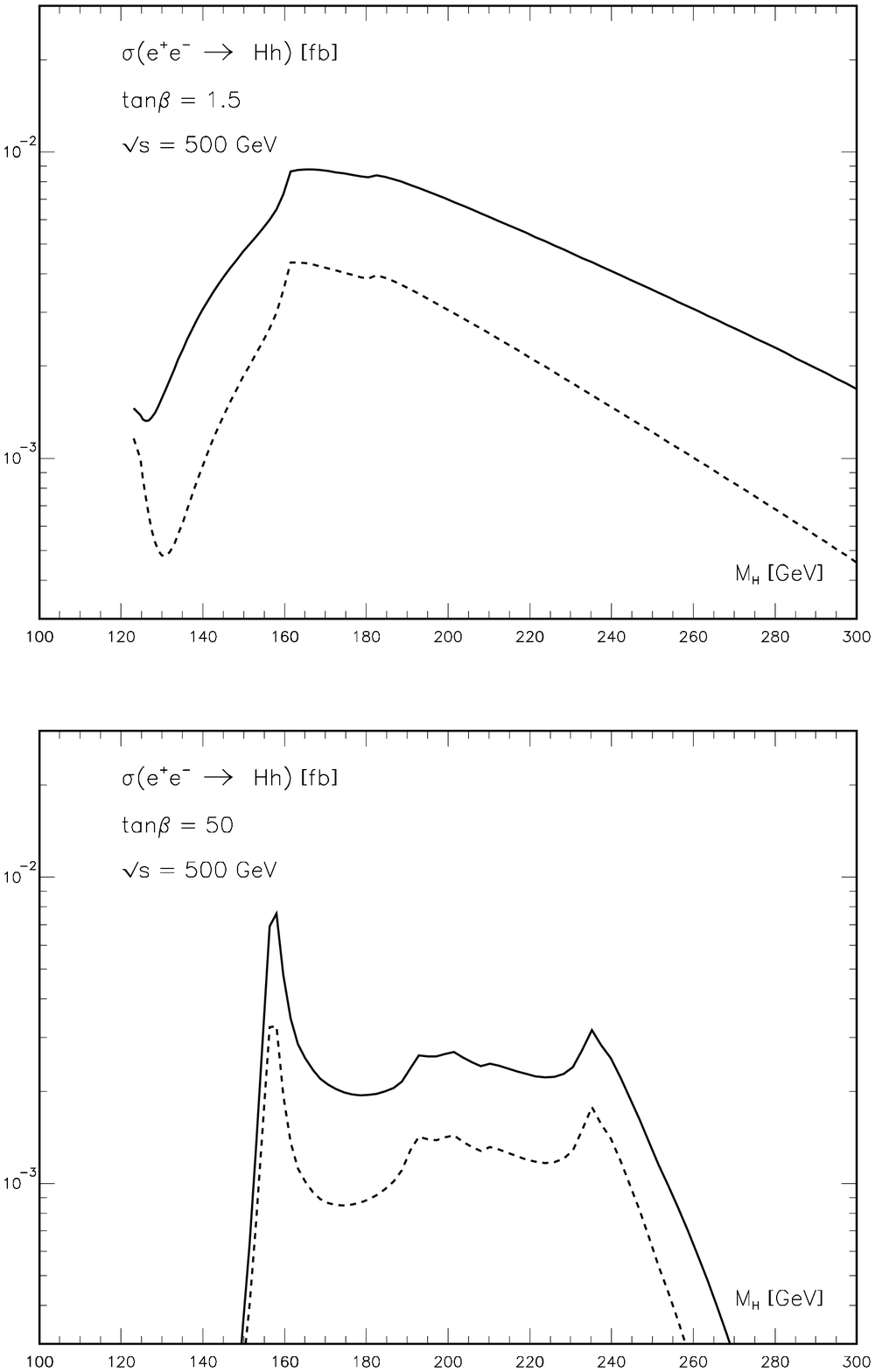,width=15cm,bbllx=30pt,bblly=50pt,bburx=550pt,bbury=760pt}
}
\end{center}                                
\centerline{\bf  Fig. 6
}\label{fig_s4}
\end{figure}                                
\end{document}